\documentclass{iopart}
\usepackage{graphicx}
\usepackage{epsfig}
\usepackage{iopams}
\usepackage{setstack}

\begin{document}

\title[Finite-size scaling of entanglement entropy]
{Finite-size scaling of the entanglement entropy of the quantum Ising chain
with homogeneous, periodically modulated and random couplings}
\author{Ferenc Igl\'oi}
\address{Research Institute for Solid
State Physics and Optics, H-1525 Budapest, P.O.Box 49, Hungary}
\address{
Institute of Theoretical Physics,
Szeged University, H-6720 Szeged, Hungary} 
\author{Yu-Cheng Lin}
\address{
Theoretische Physik, Universit\"at des Saarlandes, D-66041
Saarbr\"ucken, Germany}


\begin{abstract}
Using free-fermionic techniques we study the entanglement entropy of a block of
contiguous spins in a large finite quantum Ising chain in a transverse field,
with couplings of different types: homogeneous, periodically modulated and
random. We carry out a systematic study of finite-size effects at the quantum
critical point, and evaluate subleading corrections both for open and for
periodic boundary conditions. For a block corresponding to a half of a finite
chain, the position of the maximum of the entropy as a function of the control
parameter (e.g. the transverse field) can define the effective critical point
in the finite sample.  On the basis of homogeneous chains, we demonstrate that
the scaling behavior of the entropy near the quantum phase transition is in
agreement with the universality hypothesis, and calculate the shift of the
effective critical point, which has different scaling behaviors for open and
for periodic boundary conditions.
\end{abstract}

\maketitle

\newcommand{\bc}{\begin{center}}
\newcommand{\ec}{\end{center}}
\newcommand{\be}{\begin{equation}}
\newcommand{\ee}{\end{equation}}
\newcommand{\beqn}{\begin{eqnarray}}
\newcommand{\eeqn}{\end{eqnarray}}


\section{Introduction}

Entanglement describes nonlocal quantum correlations, and is one of the
characteristic peculiar features of quantum mechanics.  Motivated by recent
studies showing intimate connections between entanglement and quantum phase
transitions \cite{fazio1,nielsen}, the understanding of the degree of
entanglement in quantum many-body systems has prompted an enormous effort at
the interface between condensed matter physics, quantum information theory and
quantum field theory \cite{fazio}.

A fundamental question in this research field is concerned with the scaling of
the entropy quantifying the degree of entanglement between a spatially confined
region and its complement in a quantum many-body system.  Suppose a system,
combined by two subsystems $\mathcal{A}$ and $\mathcal{B}$, is in a pure quantum
state $|\psi\rangle$, with density matrix $\rho=| \psi\rangle\langle\psi |$.
The entanglement entropy is just the von Neumann entropy of either subsystem
given by
\be
S_{\cal A}=-\Tr (\rho_{\cal A} \log_2 \rho_{\cal A}) = 
           -\Tr (\rho_{\cal B} \log_2 \rho_{\cal B}) =
S_{\cal B},
\ee
where the reduced density matrix for $\cal A$  is constructed by tracing over
the degrees of freedom in $\cal B$, given by $\rho_{\cal A}=\Tr_{\cal B}\rho$.
Analogously, $\rho_{\cal B}=\Tr_{\cal A}\rho$. In order to explore the behavior
of quantum entanglement at different length scales, one is particularly
interested in how the entanglement entropy depends on the linear size $\ell$ of
the subsystem considered.  An early conjectured scaling law relates the
entanglement entropy to the surface area $\ell^{d-1}$, not the volume, of the
region in a $d$-dimensional system \cite{AREA}. This area law of entropy
scaling has been established for gapped quantum many-body systems where the
correlation length is finite.  In one-dimensional (1D) systems, the
entanglement behavior changes drastically at a quantum phase transition where
the absence of gaps leads to long range correlations and results in a
logarithmically diverging entanglement entropy as the system size $L$ goes to
infinity, i.e. $S_{\mathcal{A}}\sim \log_2 \ell$ for $L\to \infty$
\cite{vidal,conformal,calabrese_cardy,refael}. This connection between
entanglement entropy and quantum phase transitions is however lost in quantum
systems in higher dimensions \cite{FREE-F,BOSON-NUM,BOSON-ANA,lir07,yu}.

The scaling behavior of quantum entanglement for $1+1$-dimensional conformally
invariant systems has been derived by several authors
\cite{conformal,calabrese_cardy}.  Here we summarize some known results. For a
critical chain of length $L$ with  {\it periodic} boundary conditions, the
entanglement entropy of a subsystem of length $\ell$ embedded in the chain
scales as
\be
S^{(p)}_L(\ell)=\frac{c}{3} \log_2\left[\frac{L}{\pi} 
\sin\left(\frac{\ell \pi}{L}\right)\right] + c_1,
\label{S_L_p}
\ee
where the prefactor $c$ is universal and given by the central charge of the
associated conformal field theory, whereas the constant $c_1$ is non-universal.
For the leftmost segment of length $\ell$ in a finite {\it open} chain of
length $L$ at criticality, the entanglement entropy reads
\be
S^{(o)}_L(\ell)= \frac{c}{6} \log_2\left[\frac{2L}{\pi} \sin\left(\frac{\ell \pi}{L}\right)\right] +
\log_2 g +\frac{c_1}{2},
\label{S_L_o}
\ee
where $\log_2 g$ is the boundary entropy\cite{boundary_entropy} and the
constant $c_1$ is the same to the one in Eq.~(\ref{S_L_p}). For an infinite
system $L\to\infty$, the critical entanglement entropy becomes
\be
  S_\infty(\ell)=\frac{c}{3}\log_2 \ell + c_1.
  \label{S_l}
\ee 
Away from the critical point, where the correlation length $\xi\ll \ell$, we have
\be
S_\infty \simeq b\frac{c}{6} \log_2 \xi.
\label{S_xi}
\ee
where $b$ is the number of boundary points between the subsystem and the rest
of the chain.  Some of the results given above have been verified by analytic
and numerical calculations on integrable 1D quantum spin chains, in particular
on the antiferromagnetic $XX$-chain and on the quantum Ising chain
\cite{vidal,peschel05,jin_korepin,boundary}. Notice that an exact relationship
between the entanglement entropy of these two models has been recently
established \cite{IJ07}.

Remarkably, the logarithmic scaling law of entanglement entropy in the
thermodynamic limit is valid even for critical quantum chains that are not
conformally invariant. In those cases the central charge determining the
prefactor of the logarithmic scaling law is replaced by an effective one.  For
disordered quantum Ising and $XX$ chains at infinite-randomness fixed points,
the effective central charge was determined as $c_{\rm eff}=c\ln 2$ for the
disorder-average entropy \cite{refael}  by using strong disorder
renormalization group method \cite{mdh,fisher,review}. Also the average entropy
of other types of random quantum spin chains with infinite-randomness fixed
points has been studied by similar methods\cite{Santachiara,Bonesteel,s=1}. In
aperiodic quantum Ising chains, where the couplings follow some quasi-periodic
or aperiodic sequence, the coefficient in Eq.~(\ref{S_l})  is shown to depend
on the ratio of the couplings\cite{ijz07}, provided the perturbation caused by
the aperiodicity is marginal or relevant.

In this paper we consider the quantum Ising chain with three different types of
couplings: homogeneous, periodically modulated and random. We calculate the
entanglement entropy for large finite systems up to $L=4096$ by free fermionic
techniques.  For the homogeneous chain, conformal predictions about the entropy
at the critical point for finite chains with different boundary conditions are
checked, and subleading corrections are investigated. We also study the
finite-size scaling behavior of $S_L(L/2)$ around its maximum and use the
position of the maximum to identify the finite-size critical transverse field.
The model with periodically modulated couplings belongs to the same critical
universality class as the homogeneous model. In this case we study the entropy
for finite chains and check whether the logarithmic scaling law is valid.
Finally, for random chains we calculate the average entropy, check the validity
of the strong disorder renormalization group prediction, and compare the
average entropy with the corresponding conformal result in Eq.(\ref{S_L_p}).

The structure of the paper is the following. In Sec. \ref{sec:fermionic} we
present the model, its free-fermion solution and the way of calculating the
entanglement entropy. Results of the numerical calculations at the critical
point are shown in Sec. \ref{sec:crit} for homogeneous, periodically modulated
and random chains. For homogeneous chains finite-size scaling of the maximum of
the entropy close to the critical point is analyzed in Sec. \ref{sec:fss}. Our
results are discussed in Sec. \ref{sec:disc}. In \ref{sec:hom}, the correlation
matrix, which is relevant to the calculation of entanglement entropy, is
determined for the homogeneous chain at its critical point. In \ref{sec:shift}
the shift exponent of homogeneous closed chains is calculated.

\section{The quantum Ising chain and its entropy in the fermionic representation}
\label{sec:fermionic}

\subsection{The model and its free-fermion representation}

The model we consider is an Ising chain with nearest neighbor couplings $J_i$ in
a transverse field of strength $h_i$, defined by the Hamiltonian:
\be
{\cal H} =
-\frac{1}{2}\sum_{i=1}^L J_{i}\sigma_i^x \sigma_{i+1}^x-
  \frac{1}{2}\sum_{i=1}^L h_i \sigma_i^z
\label{eq:H}
\ee
in terms of the Pauli-matrices $\sigma_i^{x,z}$ at site $i$. Here we consider
three types of couplings: (i) homogeneous case with $J_i=1$ and  $h_i=h\,(>0)$;
(ii) staggered case with $J_{2i-1}=\lambda\,(>0)$, $J_{2i}=1/\lambda$, and
$h_i=h$; (iii) random case with $\{J_i\}$ and $\{h_i\}$ being independent and
identically distributed random variables.


The essential technique in the solution of ${\cal H}$ is the mapping to
spinless free fermions \cite{lsm,pfeuty}. First we express the spin operators 
$\sigma_i^{x,y,z}$ in terms of fermion creation (annihilation) operators 
$c_i^\dagger$ ($c_i$) by using the Jordan-Wigner
transformation:  $c^\dagger_i=a_i^+\exp\left[\pi i \sum_{j}^{i-1}a_j^+a_j^-\right]$
and $c_i=\exp\left[\pi i
\sum_{j}^{i-1}a_j^+a_j^-\right]a_i^-$, where $a_j^{\pm}=(\sigma_j^x \pm
i\sigma_j^y)/2$. Doing this, ${\cal H}$ can be rewritten in a quadratic form
in fermion operators:
\beqn
{\cal H}&=&
-\sum_{i=1}^{L}h_i\left( c^\dagger_i c_i-\frac{1}{2} \right) -
\frac{1}{2}\sum_{i=1}^{L-1} J_i(c^\dagger_i-c_i)(c^\dagger_{i+1}+c_{i+1})\cr
&+&\frac{1}{2}w J_L(c^\dagger_L-c_L)(c^\dagger_{1}+c_{1}).
\label{ferm_I}
\eeqn
Here the parameter $w=\exp(i\pi {\cal N}_c)$ depends on the number of fermions
${\cal N}_c=\sum_{i=1}^L c_i^\dagger c_i =1/2 \sum_{i=1}^L(1+\sigma_i^z)$, therefore one
should consider two separated sectors depending on the parity of ${\cal N}_c$. 
The ground state corresponds to the fermionic vacuum, thus  $w=1$.

In the second step, the Hamiltonian is diagonalized by a Bogoliubov transformation:
\be
\eta_k=\sum_{i=1}^L\left[ \frac{1}{2}\left(\Phi_k(i)+\Psi_k(i)\right)c_i+
   \frac{1}{2}\left(\Phi_k(i)-\Psi_k(i)\right)c_i^\dagger\right]
\ee
where the $\Phi_k(i)$ and $\Psi_k(i)$ are real and normalized: $\sum_i^L
\Phi_k^2(i)=\sum_i^L \Psi^2_k(i)=1$, so that we have
\be
\mathcal{H}=\sum_{k=1}^L \Lambda_k(\eta_k^\dagger \eta_k-1/2).
\label{free_fermion}
\ee
The fermionic excitation energies, $\Lambda_k$, and the components of the
vectors, ${\mathbf \Phi}_k$ and ${\mathbf \Psi}_k$, are obtained from the
solution of the following eigenvalue problem\cite{it}: ${\mathbf T}{\mathbf
V}_k=\Lambda_k {\mathbf V}_k$. Here ${\mathbf T}$ is a symmetric $2L \times 2L$
matrix:
\be
\mathbf{T}=\left(
\begin{array}{cccccc}
0      &  h_1   &     	&       &      & -w J_L\\
 h_1   & 0      & J_1 	&        &     &       \\
       & J_1    & 0   	& h_2    &     &       \\
       &        & \ddots & \ddots &\ddots   &   \\
       &        &    	& J_{L-1} & 0   &   h_L \\
-w J_L &        &    	&         & h_L &  0
\end{array}
\right)
\label{T}
\ee
and the eigenvectors have the components: ${\mathbf
V}_k=(-\Phi_k(1),\Psi_k(1),-\Phi_k(2),\Psi_k(2),\dots,\\-\Phi_k(L),\Psi_k(L))$.
Transforming $\Phi_k(i)$ into $- \Phi_k(i)$, $\Lambda_k$ is changed to
$-\Lambda_k$.  Thus we only restrict ourselves to the sector corresponding to
$\Lambda_k \ge 0$, $k=1,2,\dots,L$. To obtain the quantum critical point of the
system we make use of the condition that the energy of the first fermionic
excitation vanishes in the thermodynamic limit. From Eq.(\ref{T}) with $w=-1$
we obtain \cite{pfeuty,ilrm07}: 
\be
\lim_{L \to \infty} \frac{1}{L}\sum_{i=1}^L \ln J_i=\lim_{L \to \infty} \frac{1}{L} \sum_{i=1}^L \ln h_i
\label{crit_point}
\ee
Consequently, the critical point of the homogeneous chain as well as the
staggered chain is located at $h_c=1$. For the random chain the criticality
condition is given by $\overline{\ln J}=\overline{\ln h}$, where the overbar 
denotes an average over quenched disorder.

\subsection{Calculation of the entanglement entropy}

Now we turn to the procedure for calculating the  entanglement entropy of
the system in its ground state $|0\rangle$. We consider a subsystem of length
$\ell$, consisting of spins $i=1,2,\dots,\ell$. The reduced density matrix
${\mathbf \rho}_{\ell}=\Tr_{L-\ell} |0\rangle \langle 0 |$ can be
calculated from the restricted correlation matrix ${\mathbf G}$
\cite{peschel,vidal}, the elements of which are given by
\beqn
G_{m,n}=\langle 0 |(c^\dagger_n-c_n)(c^\dagger_m+c_m)|0 \rangle\cr
=-\sum_{k=1}^L \Psi_k(m) \Phi_k(n),\quad m,n=1,2,\dots,\ell
\label{G}
\eeqn
For the homogeneous critical chain, the eigenvalue problem of $\mathbf{T}$ 
in Eq.~(\ref{T}) and thus the matrix elements of ${\mathbf G}$ can be solved
analytically. The results, both for periodic and for open finite chains
with an even $L$, as well as for $L\to\infty$, are given in \ref{sec:hom}.

The von Neumann entropy of the considered subsystem,
$S_L(\ell)=-\Tr(\rho_\ell\log_2 \rho_\ell)$, is fully determined by the
spectrum of the reduced density matrix ${\mathbf \rho}_{\ell}$.  To diagonalize
${\mathbf \rho}_{\ell}$, we transform the $\ell$ fermionic modes into
non-correlated fermions with operators:
\be
\mu_q=\sum_{i=1}^\ell\left[ \frac{1}{2}\left(v_q(i)+u_q(i)\right)c_i+
   \frac{1}{2}\left(v_q(i)-u_q(i)\right)c_i^\dagger\right]\;,
\label{mu_q}
\ee
where the $v_q(i)$ and $u_q(i)$ are real and normalized: $\sum_i^{\ell} v_q^2(i)=\sum_i^{\ell} u_q^2(i)=1$.
In the transformed basis, we have
\be
\langle 0 |\mu_q \mu_p|0 \rangle=0,\quad \langle 0 |\mu_q^\dagger \mu_p|0 \rangle=\delta_{qp}\frac{1+\nu_q}{2}\;,
\label{uncorr}
\ee
for $p,q=1,2,\dots \ell$, which means that the fermionic modes are uncorrelated.
Thus the reduced density matrix is the direct
product  $\rho_\ell=\bigotimes_{q=1}^{\ell} \rho_q$, where $\rho_q$ has eigenvalues
$(1 \pm \nu_q)/2$. The entanglement
entropy is then given by the sum of binary entropies:
\be
S_L(\ell)=-\sum_{q=1}^{\ell} \left(\frac{1+\nu_q}{2} \log_2 \frac{1+\nu_q}{2}
+\frac{1-\nu_q}{2} \log_2 \frac{1-\nu_q}{2}\right).
\label{binary}
\ee
The $\nu_q$-s in Eq.~(\ref{binary}) are the solutions of the equations
\be
{\mathbf G}{\mathbf u}_q=\nu_q {\mathbf v}_q,\quad
{\mathbf G}^T{\mathbf v}_q=\nu_q {\mathbf u}_q\;,
\ee
or, equivalently, are related to the eigenvalue problem:
\be
{\mathbf G}{\mathbf G}^T{\mathbf v}_q=\nu_q^2 {\mathbf v}_q,\quad
{\mathbf G}^T{\mathbf G}{\mathbf u}_q=\nu_q^2 {\mathbf u}_q\;.
\label{G_GT}
\ee
In numerical calculations, many eigenvalues $\nu_q^2$ are found to be very
close to zero and these small eigenvalues are often out of the computer
precision, resulting in instability in the calculations.  To circumvent the
problem, we can introduce a symmetric $2\ell \times 2\ell$ matrix ${\mathbf U}$
with elements:
\be
U_{i,j}=
\left[\begin{array}{cc}
0                   & G_{i,j}         \cr
G_{j,i}             & 0               
\end{array}\right]\;,
\ee
whose eigenvalue problem corresponds to ${\mathbf U}{\mathbf W}_q=\nu_q {\mathbf W}_q$. 
Here the eigenvector ${\mathbf W}_q$ is given by
${\mathbf W}_q=\left(-v_q(1),u_q(1),-v_q(2),u_q(2),\dots,-v_q(\ell),u_q(\ell) \right)$.
Only non-negative eigenvalues $\nu_q \ge 0$ are taken into account.

\section{Scaling at the critical point}
\label{sec:crit}
Here we calculate the entanglement entropy of the quantum Ising chain for different types of
interactions (homogeneous, staggered and random) at the quantum critical point, defined
in Eq.(\ref{crit_point}). The numerical results obtained for finite periodic and open
chains are compared with the conformal results in Eqs.(\ref{S_L_p}) and (\ref{S_L_o}),
respectively.

\subsection{Homogeneous chain}
\label{sec:crit_hom}

For a critical Ising chain of finite length $L$ with periodic boundary
conditions, the expression for the entanglement entropy of a subsystem of size
$\ell$ is given by Eq.~(\ref{S_L_p}) with the central charge $c=1/2$, while
with open boundary conditions it corresponds to Eq.~(\ref{S_L_o}) with the
theoretical value $g=1$ \cite{cardy_g} for the boundary entropy $\log_2 g$.  To
calculate the non-universal constant $c_1$ we use the exact relationship:
$S_{2L}^{XX}(2\ell)=2S_L(\ell)$ \cite{IJ07}, between the entropy of the
$XX$-chain, $S_L^{XX}(\ell)$, and the entropy of the quantum Ising chain,
yielding $c_1=\frac{1}{2}c_1^{XX}+\frac{1}{3} c$, where $c_1^{XX}$ is the
constant for the $XX$-chain and is given in Ref.~\cite{jin_korepin} in terms of
a definite integral. In this way we obtain $c_1=0.6904132738\cdots$, which
agrees with the value evaluated in a recent paper using a different method
\cite{cardy}. In Table \ref{table:check}, the constant $c_1$ is calculated by
\be
c_1(L)=S_L^{(p)}(\ell)-\frac{c}{3} \log_2\left[\frac{L}{\pi} \sin\left(\frac{\ell \pi}{L}\right)\right],
\label{c_1}
\ee
with $\ell=L/2$. It, indeed, converges to the asymptotic value of $c_1$ as the system
size $L$ is increasing.  
\begin{table}
\caption{Finite-size dependence of the ratio $r(L)$ defined in Eq.~(\ref{ratio}) for
chains with periodic boundary conditions, the constant $c_1(L)$ in Eq.(\ref{c_1}) with
$\ell/L=1/2$, and $g$ for the boundary entropy $\log_2 g$ in Eq.(\ref{S_L_o}).
The finite-size correction coefficients are: 
$r_2=0.3334(2)$ and $g_1=0.3095(2)$\label{table:check}
}
\begin{indented}
\item[]\begin{tabular}{|c|c|c|c|}  \hline
  L   &   $r$        & $c_1$ & $g$  \\ \hline
  128 & 0.5000203756 & 0.6904174985 & 0.9975945594  \\
  256 & 0.5000050926 & 0.6904143299 & 0.9987944047  \\
  512 & 0.5000012731 & 0.6904135378 & 0.9993964859  \\
 1024 & 0.5000003182 & 0.6904133397 & 0.9996980631  \\
 2048 & 0.5000000795 & 0.6904132903 & 0.9998489873  \\
 4096 &              & 0.6904132780 & 0.9999244833  \\ \hline
$\infty$ & $0.5+r_2/L^{2}$ & 0.6904132738 & $1-g_1/L$ \\ \hline
  \end{tabular}
\end{indented}
  \end{table}

\begin{figure}[t]
\begin{center}
\includegraphics[width=3.2in,angle=0]{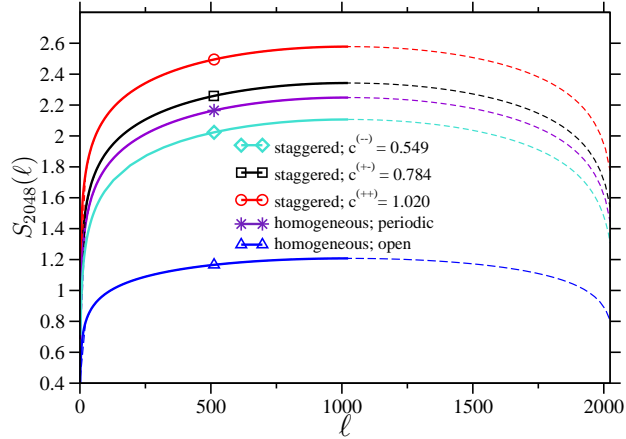}
\end{center}
\caption{
\label{fig:profile}
Entropy of finite chains of length $L=2048$ vs subsystem size $\ell$: the full
curves for $\ell \le L/2$ are calculated numerically; the dashed curves for
$\ell \ge L/2$ are the corresponding conformal results, described by
Eq.~(\ref{S_L_p}) with $c=1/2$ and $c_1=0.690413$ for homogeneous chains with
periodic boundary conditions, and by Eq.~(\ref{S_L_o}) with $g=1$ for open
homogeneous chains.  For chains with staggered interactions, there are four
branches, depending on the type of the couplings on the boundary of the
subchain. Data presented here are for staggered chains with $\lambda=0.5$ and
with periodic boundary conditions.  The fits for $\ell \ge L/2$ using
Eq.~(\ref{S_L_p}) are fulfilled with $c=1/2$, $c_1^{(++)}=1.02009$ (two strong
couplings on the boundary), $c_1^{(+-)}=c_1^{(-+)}=0.78446$ (one strong and one
weak coupling) and $c_1^{(--)}=0.54883$ (two weak couplings), respectively.}   
\end{figure}

A comparison between the conformal expressions (in Eq.~(\ref{S_L_p}) and
Eq.~(\ref{S_L_o})) and the entropy calculated by exact diagonalization for
$L=2048$ is shown in Fig.~\ref{fig:profile}, both with periodic and open
boundary conditions; an excellent agreement is achieved. The 
accuracy of the functional form can be checked by the ratio
\be
r(L)=\frac{S_L(L/2)-S_L(L/4)}{S_{2L}(L)-S_{L}(L/2)}\quad \underset{L\to\infty}{\rightarrow}\quad \frac{1}{2}\;.
\label{ratio}
\ee
The numerical results for $r(L)$ for different system sizes $L$ up to $L=2048$ are given
in Table \ref{table:check}, and the first correction term is found to be $O(L^{-2})$.


Furthermore, we are interested in the finite-size correction terms for the
coefficient $c$ and the boundary entropy $\log_2 g$ given in Eq.~(\ref{S_L_o}).
To evaluate $c$ for different system size, we first calculate the entropy
difference $\Delta S(L)=S_L(L/2)-S_{L/2}(L/4)$. For a chain with periodic
boundary conditions, we have $\Delta S(L)=c(L)/3$ and obtain an $L^{-2}$ -
correction for the coefficient: $c(L)=0.5-0.623(1)/L^2+O(L^{-3})$; for an open
chain, we obtain a $L^{-1}$ - correction, $c(L)=0.5+1.339(1)/L+O(L^{-2})$, via
$\Delta S(L)=c(L)/6$. To compute the boundary entropy $g(L)$, we make use the
relation between  $S_L^{(p)}$ and $S_L^{(o)}$ for periodic and open boundary
conditions, respectively, via $2S_L^{(o)}(\ell)-S_L^{(p)}(\ell)=\log_2
g(L)+c/3$. In Table \ref{table:check}, the values of $g(L)$ using $\ell=L/2$
and $c=1/2$ are given for system sizes up to $L=4096$, and it shows a
correction of $O(1/L)$.

In conclusion, our numerical results for the entanglement entropy of
the homogeneous quantum Ising chain agree with all the known conformal predictions.

\subsection{Chains with staggered interactions}
\label{sec:stagg}
Now we consider the quantum Ising chain with periodically varying interactions
of period 2, corresponding to a chain with staggered interactions:
$J_{2i-1}=\lambda$ and $J_{2i}=1/\lambda$.  According to
Eq.~(\ref{crit_point}), the critical point of the system is located at $h_c=1$.
This quantum Ising chain with staggered interactions has been solved in
Ref.~\cite{iz88}, and its critical singularities were found to be the same as for
the homogeneous chain.  Here we study the entanglement entropy of the staggered
chain and check its relationship with the entropy of the homogeneous chain.

First, we calculate the entanglement entropy, $S_L(\ell)$, as a function of the
subsystem  size $\ell$, for a finite chain. As shown in
Fig.~\ref{fig:profile} for a chain of length $L=2048$ with  $\lambda=0.5$,
there are four branches with a twofold degeneracy, depending on the type of the
couplings ($\lambda$ or $\lambda^{-1}$) at the boundaries of the subsystem. For
each $\ell$, the largest and smallest value of $S_L(\ell)$ correspond to the
case in which both boundary couplings are strong (denoted by $(++)$) and weak
($(--)$), respectively, and the twofold degeneracy lying in between occurs when
one boundary coupling is strong and one weak ($(+-)$ and $(-+)$). All branches
are well fitted by the conformal form in Eq.~(\ref{S_L_p}) with coefficient
$c=1/2$ corresponding to the central charge of the homogeneous case, but with
different additive constants $c_1$. The additive constants for the above mentioned
four branches satisfy the relation: $c_1^{(++)}+c_1^{(--)}=2c_1^{(+-)}$.
This means that the boundary effect is strictly additive:
$c_1^{(++)}=2c_1^{(+)}$, $c_1^{(+-)}=c_1^{(+)}+c_1^{(-)}$ and
$c_1^{(--)}=2c_1^{(-)}$, here the subscript $+$ ($-$) corresponds to one strong
(weak) boundary coupling.  For $\lambda=0.5$ we have $c_1^{(+)}=0.510045$ and
$c_1^{(-)}=0.274415$, whereas for $\lambda=0.25$ these are $c_1^{(+)}=0.663435$
and $c_1^{(-)}=0.262845$.  Furthermore, $c_1^{(+)}$ ($c_1^{(-)}$) is found to
be a monotonously increasing (decreasing) function of $1/\lambda \ge 1$, and the
average, $c_1^{(+-)}/2$, is minimal for the homogeneous chain $\lambda=1$.
Consequently, for irrelevant perturbations represented by the staggered
interaction the average critical entanglement entropy is increasing, compared
with the fixed point value of the homogeneous chain.

To see how the coefficient $c(L)$ for a finite chain of length $L$ approaches
the conformal value $c=1/2$, we follow the procedure described in
Sec.~\ref{sec:crit_hom} for the homogeneous chain. Like the homogeneous chain,
the leading term of the finite-size correction to $c(L)$ is found to be
$O(L^{-2})$ for periodic boundary conditions, and $O(L^{-1})$ for
open boundary conditions. 

\subsection{Random chains}

The entanglement entropy of the quantum Ising model with random couplings
and/or transverse fields can be conveniently studied by the strong disorder
renormalization group (RG) method\cite{refael,review}. In this RG
representation, the ground state of the quantum Ising model consists of a
collection of independent ferromagnetic clusters of various sizes; each cluster
of $n$ spins is in a $n$-site entangled state
$\frac{1}{\sqrt{2}}(\left|\uparrow\right\rangle^{\otimes
n}+\left|\downarrow\right\rangle^{\otimes n})$.  The entanglement entropy of a subsystem
is just given by the number of the clusters that cross the boundary of the
subsystem.  In 1D the asymptotic number of such clusters that contribute
to the entropy of a subsystem of length $\ell$ has been analytically calculated 
by Refael and Moore \cite{refael} and the disorder average entropy in the long chain limit 
is found to scale as:
\be
\overline{S}(\ell)=\frac{c_{\rm eff}}{3} \log_2 \ell + c_1'
\label{S_l_r}
\ee
where the effective central charge, $c_{\rm eff}=\ln 2/2$, is expected to be universal,
i.e. does not depend on the form of disorder, whereas the additive constant, $c_1'$,
is disorder dependent. 

For a finite chain of length $L$ with periodic boundary conditions, the entropy of
a subsystem of length $\ell$ is expected to behave as:
\be
\overline{S}_L(\ell)=\frac{c_{\rm eff}}{3} \log_2[ L f(\ell/L)]+ c_1'
\label{S_c_eff}
\ee
where the scaling function $f(v)$ is reflection symmetric, $f(v)=f(1-v)$, and
$\lim_{v \to 0} f(v) \simeq v$. Consequently $f(v)$ can be expanded as a
Fourier series: $f(v)=\sum_{k=1}^{\infty} A_k \sin (2k-1) \pi v$, with
$\sum_{k=1}^{\infty} A_k (2k-1) \pi=1$. We note that for 
conformally invariant models only the first term of this expansion exists
(cf. Eq.~(\ref{S_L_p})).

In our numerical calculations we used a power-law distribution:
\be
P_D(x)=\frac{1}{D} x^{-1+1/D}\;,
\ee
both for the couplings and the transverse fields, which ensures that the random
model is at the critical point. Here $D^2=\overline{\ln^2x}-\overline{\ln x}^2$
measures the strength of disorder.  For the random chains we have treated
finite chains up to a length $L=1024$, and considered at least $10^4$
independent realizations for each length $L$, plus different positions of
a subsystem in the chain for a given $\ell$. 


\begin{figure}[t]
\begin{center}
\includegraphics[width=3.2in,angle=0]{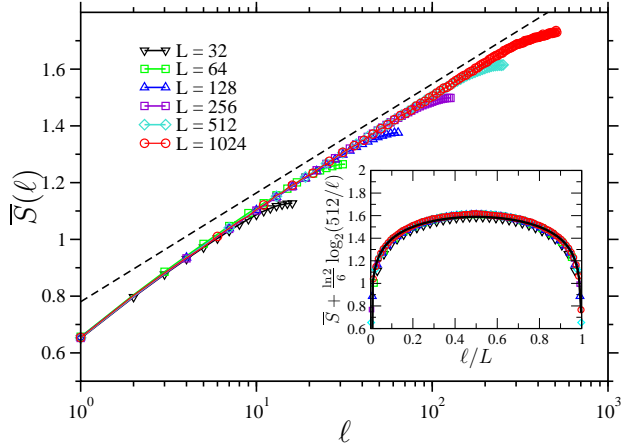}
\end{center}
\caption{
\label{fig:random}
Average entropy of the random chain with uniform disorder (with $D=1$) as a
function of $\ln \ell$ for different system sizes $L$. The slope of the broken
straight line is given by $1/6$, corresponding to the RG prediction $c_{\rm
eff}=\ln 2/2$. Inset: scaling plot of the entropy vs. $\ell$ for different sizes
$L$ using the scaling prediction in Eq.~(\ref{S_c_eff}). The solid line corresponds
to the conjecture of conformal invariance, given in Eq.~(\ref{S_L_p}), 
with $c$ replaced by $c_{\rm eff}=\ln(2)/2$ and an additive constant as the fit parameter.  
}
\end{figure}


In Fig.~\ref{fig:random} we plot the average entropy $\overline{S}_L(\ell)$ vs.
$\ln \ell$.  The curves tend to approach an asymptotic linear behavior with a
slope which is, in the large $\ell$ regime, well described by the
renormalization group prediction $c_{\rm{eff}}=\ln(2)/2$. To estimate
$c_{\rm eff}(L)$ quantitatively for different chain sizes $L$, we average over
the entropy in the large $\ell$ region and make use of the relation between
$\overline{S}_{2L}$ and $\overline{S}_{L}$, given by
\be
\frac{1}{2n+1} \sum_{\ell=L/2-n}^{L/2+n}[\overline{S}_{2L}(2l)-\overline{S}_L(l)]=
c_{\rm eff}(2L)/3.
\label{c_L_n}
\ee
The estimated values of $c_{\rm eff}(L)/\ln 2$ are presented in Table
\ref{table:random} for uniform disorder. The results for the two largest finite
systems are compatible with the estimate: $c_{\rm eff}/\ln 2=0.501(3)$, which is
in excellent agreement with the RG prediction.
\begin{table}
\caption{Finite-size estimates of $c_{\rm eff}/\ln 2$ for the random model
using the relation in Eq.(\ref{c_L_n}). \label{table:random}\\}
\begin{indented}
\item[]\begin{tabular}{|c|c|c|c|}  \hline
  n   & $L=128$ & $L=256$ &  $L=512$ \\ \hline
  2 & 0.531 & 0.503 & 0.502  \\
  4 & 0.532 & 0.502 & 0.506 \\
  8 & 0.533 & 0.502 & 0.501  \\
 16 & 0.534 & 0.504 & 0.499  \\  \hline
  \end{tabular}
\end{indented}
  \end{table}

Finally we turn to a study of the form of the average entropy
$\overline{S}_L(\ell)$ as a function of $\ell$, in particular we are interested
in how well it can be approximated by the conjecture of conformal invariance
given in Eq.~(\ref {S_L_p}) with an effective central charge
$c_{\rm{eff}}$.  As shown in the inset of Fig.~\ref{fig:random}, the
approximation is seemingly good.  To have a quantitative comparison, we have
calculated the ratio $\overline{r}(L)$, similar to Eq.~(\ref {ratio}), defined
as
\be
\overline{r}(L)=\frac{\overline{S}_L(L/2)-
                \overline{S}_L(L/4)}{\overline{S}_{2L}(L)-\overline{S}_{L}(L/2)},
\label{ratio_r}
\ee
whose asymptotic value is given, in terms of the
Fourier coefficients, by:
\beqn
\overline{r}&=&\frac{1}{2} + \log_2 \left[ \sum_{k=1}^{\infty} (-1)^{k+1}A_k \right]\cr
&-&\log_2 \left[ \sum_{k=1}^{\infty} (-1)^{k+1}(A_{2k-1}+A_{2k}) \right] \;.
\eeqn
For conformal invariant cases, we have $\overline{r}=1/2$.  The numerically calculated
values of $\overline{r}(L)$, presented in Table~\ref{table:rand} for disorder
strength $D=0.5$ and $D=1$, deviate significantly from $\overline{r}=1/2$ for large $L$. 
This means that the higher order terms in the Fourier expansion are not
negligible. The scaling function $f(v)$ is presumably universal, i.e.
independent of the form of the disorder. 
\begin{table}
\caption{The ratio defined in Eq.~(\ref {ratio_r}) for the average entropy $\overline{S}_L(\ell)$
for disorder strength $D=0.5$ and $D=1$.\label{table:rand}\\}
\begin{indented}
\item[]
\begin{tabular}{|c|c|c|}  \hline
  L   &   $D=0.5$        & $D=1.0$ \\ \hline
  32 &  0.548 &  0.587   \\
  64 &  0.588 &  0.605   \\
 128 &  0.573 &  0.584   \\
 256 &  0.619 &  0.608   \\
 512 &  0.615 &  0.606   \\ \hline
 
  \end{tabular}
\end{indented}
  \end{table}

\section{Scaling close to the critical point}
\label{sec:fss}

So far we have studied the entanglement entropy at the critical point.  In this
section, we consider the entanglement between two halves of a finite chain with
homogeneous interactions, and study its behavior approaching to the critical point
$h_c=1$. 


\begin{figure}
\begin{center}
\includegraphics[width=3.2in,angle=0]{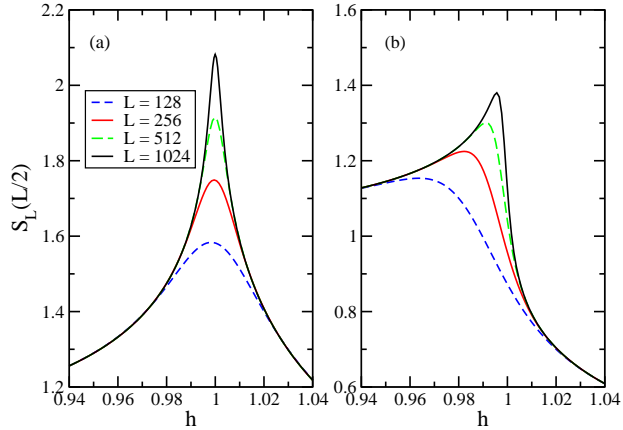}
\end{center}
\caption{
\label{fig:S_h} 
The entanglement entropy
of a half of a chain with $J=1$ as a function of the strength of the
transverse field $h$, for periodic (a)
and open (b) boundary conditions. On increasing the system size
$L$, the maximum gets more pronounced, and the position of the
maximum tends towards the critical point $h_c=1$.  
}
\end{figure}

\begin{figure}
\begin{center}
\includegraphics[width=3.2in,angle=0]{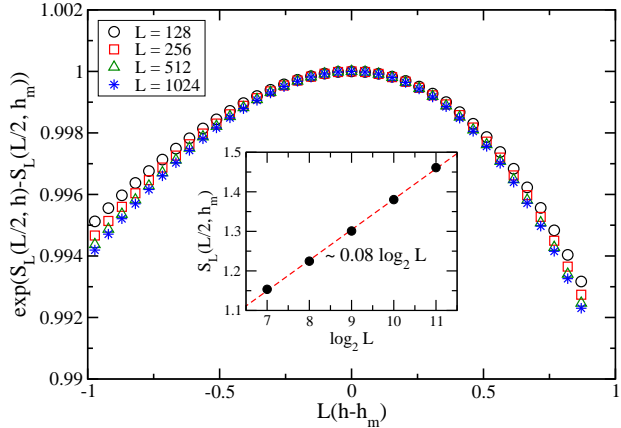}
\end{center}
\caption{
\label{fig:S_sc_o} 
A scaling plot of the entanglement entropy
for chains with open boundary conditions. Data collapse is obtained for
$\nu=1$, consistent with the universality hypothesis. In the inset is shown
the divergence of the value at the maximum as the system size increases. The slope
is given by 0.08, consistent with the exact value $1/12$ (cf. Eq.~(\ref{S_xi})).} 
\end{figure}
\begin{figure}
\begin{center}
\includegraphics[width=3.2in,angle=0]{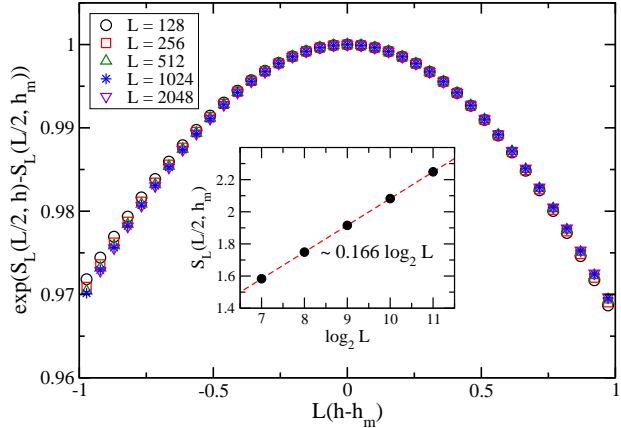}
\end{center}
\caption{
\label{fig:S_sc_p} 
The finite size scaling is performed for
chains with periodic boundary conditions, using the scaling form in 
Eq.~(\ref{diff_S1}) and $\nu=1$. The inset shows the logarithmic dependence
of the value at the maximum on the system size $L$. The slope is consistent with 
the theoretical value $1/6$ for a half subchain taken from a closed chain.}
\end{figure}

According to Eq.~(\ref{S_l}) and Eq.~(\ref{S_xi}), a divergence of the maximal
entanglement entropy occurs at the quantum critical point, which can be traced
back to the divergence of the correlation length with $\xi\sim |h-h_c|^{-\nu}$.
In a finite system of length $L$, the finite size effects induce a rounding and
a shift of the maximum of the entropy, as shown in Fig.~\ref{fig:S_h} for
$S_L(L/2)$ vs. $h$.  In the following we denote the entropy of a half of a
finite chain of length $L$ as a function of the transverse field $h$ by
$S_L(L/2,h) \equiv \hat{\cal S}(L,h)$. The position of the maximum of
$\hat{\cal S}(L,h)$, denoted by $h_m(L)$, can be used to define a finite-size
effective critical point and its shift from the true critical point is expected
to scale as: $h_c-h_m(L) \sim L^{-\lambda}$, where $\lambda$ is the shift
exponent. The numerically calculated finite-size transition points are listed
in Table~\ref{table:fss} both for closed and open chains. The maximum of the
entropy $\hat{\cal S}(L,h_m(L))$, like $\hat{\cal S}(L,h_c)$, depends
logarithmically on the system sizes $L$ [insets in Fig.~\ref{fig:S_sc_o} and
Fig.~\ref{fig:S_sc_p}].  
As a matter of fact the difference $\Delta S(L)=\hat{\cal S}(L,h_m(L))- \hat{\cal
S}(L,h_c)$ approaches a well defined limiting value for $L \to \infty$.  Note,
however that for open chains $\Delta S(L)$ tends to a finite value, whereas for
closed chains the entropy difference goes to zero.

We first study the rounding of the maximum of the entropy. Making
use of the finite-size scaling ansatz\cite{barber}:
\be 
\hat{\cal S}(L,h)-\hat{\cal S}(L,h_m(L))= \widetilde{F}[L^{1/\nu}(h-h_m(L))]\;.
\label{diff_S1} 
\ee 
with $\nu=1$, we can make all data for different system sizes perfectly
collapse onto a single curve, as shown in Fig.~\ref{fig:S_sc_o} for open chains
and Fig.~\ref{fig:S_sc_p} for closed chains. In both cases the scaling function
is $\widetilde{F}[\tilde{x}] \sim \tilde{x}^2$ for small $\tilde{x}$.

In order to obtain the shift of the finite-size critical points, we take the
derivative of both sides of Eq.(\ref{diff_S1}) at $h=h_c$:
\be
\left.\frac{\partial \hat{\cal S}(L,h)}{\partial h}\right|_{h_c} \sim L^{2/\nu}(h_c-h_m(L))\;.
\label{der_S}
\ee
For open chains the derivative at the l.h.s. is proportional to $L^{1/\nu}$,
leading to conventional finite-size scaling relation: $h_c-h_m(L) \sim
L^{-1/\nu}$. For closed chains this derivative has a much weaker
$L$-dependence, which can be identified as logarithmic in $L$.  From
Eq.~(\ref{der_S}), we then expect the relation: $h_c-h_m(L) \sim \log_2 L/L^2$.
This prediction can be checked by calculating the shift exponent $\lambda$
through the finite-size estimates:
$\lambda(L)=\log_2[h_c-h_m(L/2)]-\log_2[h_c-h_m(L)]$ [Table \ref{table:fss}].
For open chains the exponent approaches $\lambda=1/\nu=1$ for large $L$, in
accordance with our previous discussion. For closed chains the effective shift
exponent is around $1.87$ for the largest system size, which, however, cannot
rule out a true value $\lambda=2$ with a logarithmic finite-size-correction.
In \ref{sec:shift} we  present an argument in favor of the
$\log_2 L/L^2$ behavior of the shift for closed chains.
As a numerical check of this scenario we have calculated
the scaling combination: $sc =(1-h_m(L))\times L^2/(\log_2 L+a)$,
which are shown in Table \ref{table:fss}. Indeed the value of $sc$
seems to approach a finite limiting value.


Having clarified the finite-size scaling behavior of the rounding and the shift
of the maximum of the entropy, let us consider the scaling form of the entropy
in the critical region. In the conventional finite-size scaling theory we have the
ansatz:
\be
\hat{\cal S}(L,h)-\hat{\cal S}(L,h_c)= F[L^{1/\nu}(h-h_c)]\;.
\label{diff_S2} 
\ee 
The l.h.s. of Eq.(\ref{diff_S2}) can be rewritten as:
$\hat{\cal S}(L,h)-\hat{\cal S}(L,h_m(L))+\Delta S(L)=\widetilde{F}[\tilde{x}]+\Delta S(L)$,
where
the argument of $\widetilde{F}(\tilde{x})$ is given by: $\tilde{x}=x-x_c$ with
$x=L^{1/\nu}(h-h_c)$ and $x_c=L^{1/\nu}(h_c-h_m(L))$. Now using the fact
that $\widetilde{F}[\tilde{x}]$ is quadratic for small $\tilde{x}$ we obtain
for the scaling function in Eq.(\ref{diff_S2}):
\be
F(x)\approx A(x-x_c)^2+\Delta S, \quad x \approx x_c\;.
\ee
For open chains in the large $L$ limit, we have $x_c>0$ and $\Delta S>0$, 
so that shift exponent is $\lambda=1/\nu$. On the other hand, for closed chains
both limiting values vanish: $x_c=0$ and $\Delta S=0$. Therefore
conventional finite-size scaling is not valid and the shift
exponent is $\lambda > 1/\nu$. 

We close this section by two remarks. First we note that in Ref~\cite{fazio1}
the derivatives of the nearest-neighbor concurrence, $\partial_J C(1)$, with
respect to the control parameter $J$ is studied. The position of the minimum of
$\partial_J C(1)$, which defines the effective quantum critical point of a
finite closed chain of length $L$, is shifted from the true critical point by
$L^{-1.87}$ (see Fig. 1, in Ref.~\cite{fazio1}), with a shift exponent that is
very close to the effective exponent given in Table \ref{table:fss}. One might
think that the shift of the minimum of $\partial_J C(1)$ has the same scaling
behavior as discussed here for the position of the maximum of the entropy.

Our second remark concerns random chains. The position of the maximum of the
average entanglement entropy of a half chain can be used to define sample
dependent pseudo\-critical point. Its scaling has been studied in detail in 
Ref.~\cite{ilrm07}.

\begin{table}
\caption{Finite-size critical transverse fields of the homogeneous quantum Ising
chain of $L$ sites with periodic [p] and open [o] boundary conditions calculated
from the location of the maxima of the entropy for $\ell=L/2$. The effective shift
exponents, $\lambda(L)$, are calculated by two point fits. The scaling
combinations are: $sc=(1-h_m(L))\times L^2/(\log_2 L + a)$ with $a=2-1/\ln 2$ 
(s. Appendix~\ref{sec:shift}), for periodic chains and $x_c=(1-h_m(L)) \times L$ for 
open chains.\label{table:fss}}
\begin{indented}
\item[]
 \begin{tabular}{|c||c|c|c||c|c|c|}  \hline
  L & $h_m(L)[p]$ & $\lambda(L)$&$sc$& $h_m(L)[o]$ &$\lambda(L)$& $x_c$ \\ \hline
  128 & 0.9983031 & 1.813 & 3.679 & 0.9636656 & 1.077 & 4.651 \\
  256 & 0.9995225 & 1.829 & 3.656 & 0.9822266 & 1.045 & 4.550 \\
  512 & 0.9998671 & 1.845 & 3.644 & 0.9912353 & 1.027 & 4.487 \\
 1024 & 0.9999633 & 1.858 & 3.641 & 0.9956543 & 1.015 & 4.450 \\
 2048 & 0.9999900 & 1.876 & 3.629 & 0.9978379 & 1.007 & 4.424 \\ \hline
  \end{tabular}
\end{indented}
  \end{table}

\section{Discussion}
\label{sec:disc}

In this paper we have studied finite size effects of entanglement entropy of
the quantum Ising chain at/near its order-disorder quantum phase transition.
The model considered can be expressed in terms of free fermions, which enables
us to perform large scale numerical investigations. Three types of couplings
were considered: homogeneous, periodically modulated and random couplings.   

For the homogeneous system at the critical point we have verified the finite
size form predicted by the conformal field theory, both for periodic and open
boundary conditions. We have also calculated the additive constant to the
entropy and subleading corrections. In the off-critical region, we have studied
the finite-size scaling behavior of the entropy, $S_L(L/2)$, in the vicinity of
its maximum, and confirmed the intimate connection between entanglement and
universality. The position of the maximum, $h_m(L)$, can be regarded as an
indicator of the effective critical point in the finite sample. For an open
chain the shift of $h_m(L)$ from the true critical point is shown to be
$O(L^{-1})$, whereas for a chain with periodic boundary conditions it is
$O(L^{-2} \ln L)$. We have provided analytical results for the off-critical entropy
in infinite chains to explain these findings. We expect that the shift of $h_m(L)$
for other critical quantum spin chain has the same type differences for open
and periodic boundary conditions.

The quantum Ising chain with periodically modulated couplings belongs to the
same critical universality class as the homogeneous model. In the case of
staggered couplings, we have found that the critical entropy is split into four
branches, each of which has the same prefactor (the central charge) of the
logarithm but has different additive constants.  This is expected to be generic
to critical quantum spin chains with all kinds of periodically modulated
couplings.

For random quantum Ising chains, we have numerically verified the prefactor of
the logarithm predicted by the analysis of the strong disorder renormalization
group. The functional form of the average entropy versus subsystem size, which
is presumably universal for any strength of disorder, has been found to
deviate from the results for conformally invariant models.  

The results obtained in this paper, though only based on quantum Ising chains,
are expected to be valid in some other cases of quantum spin chains.  For
example, the XY-chain is related with the Ising chain via an exact mapping, so
that the results obtained for the Ising chain can be directly transferred to
those for the XY-chain through the mapping. This mapping is also applicable for
random cases. Moreover, for random cases the criticality of many quantum spin
chains belongs to the same universality class (cf. random XX-chains and random
Heisenberg chains), known from the strong-disorder renormalization group, 
and the universality of the associated effective central charge was numerically
confirmed \cite{num}. Therefore, our results for the random cases,
e.g. the functional form of the average entropy vs. $\ell$, should be universal
for a wide range of models. 

\ack{We thank R. Juh\'asz, C. Monthus, H. Rieger and Z. Zimbor\'as for useful
discussions. This work has been supported by the National Office of Research
and Technology under Grant No. ASEP1111, by German-Hungarian exchange programs
(DAAD-M\"OB and DFG-MTA) and by the Hungarian National Research Fund under
grant No OTKA TO48721, K62588, MO45596 and M36803.}

\appendix

\section{The correlation matrix for homogeneous chains}
\label{sec:hom}

For the critical homogeneous chain the free-fermion transformation can be performed
analytically, both for closed and for open finite chains. In the following we consider
the case where the length of the chain $L$ is even.

\subsection{Closed chain}

For a closed chain with $J_L=J=1$
and $h=1$ the positive eigenvalues of Eq.(\ref{T}) are two-fold degenerate,
which are given by:
\be
\Lambda_k=2\sin \left[ \frac{2k-1}{L}\frac{\pi}{2}\right]\;,
\ee
for $k=1,2,\dots,L/2$.
One set of the eigenvectors is:
\beqn
\phi^{(1)}_k(j)&=&(-1)^j \sqrt{\frac{2}{L}} \sin \left[ \frac{2k-1}{L}\left(j-1/2\right)\pi\right]\cr
\psi^{(1)}_k(j)&=&(-1)^{j} \sqrt{\frac{2}{L}} \cos \left[ \frac{2k-1}{L}j\pi\right]\;,
\label{phi_psi_p1}
\eeqn
and the second set is:
\beqn
\phi^{(2)}_k(j)&=&(-1)^{j+1} \sqrt{\frac{2}{L}} \cos \left[ \frac{2k-1}{L}\left(j-1/2\right)\pi\right]\cr
\psi^{(2)}_k(j)&=&(-1)^{j} \sqrt{\frac{2}{L}} \sin \left[ \frac{2k-1}{L}j\pi\right]\;,
\label{phi_psi_p2}
\eeqn
Then the reduced correlation matrix is given by:
\be
G_{m,n}= \frac{(-1)^{m-n}}{L
\sin\left(\frac{\pi}{2L}[2(m-n)+1]\right)}.
\ee

\subsection{Open chain}

For open chains with $J_L=0$ the solution at the critical point reads:
\beqn
\phi_k(j)&=&(-1)^j \frac{2}{\sqrt{2L+1}} \cos \left[ \frac{2k-1}{2L+1}\left(j-1/2\right)\pi\right]\cr
\psi_k(j)&=&(-1)^{j+1} \frac{2}{\sqrt{2L+1}} \sin \left[ \frac{2k-1}{2L+1}j\pi\right]\;,
\label{phi_psi_o}
\eeqn
and the energy of the free-fermionic modes are given by:
\be
\Lambda_k=2\sin \left[ \frac{2k-1}{2L+1}\frac{\pi}{2}\right]\;.
\label{lambda_o}
\ee
for $k=1,2,\dots L$.
The reduced correlation matrix reads:
\be
G_{m,n}= \frac{(-1)^{m+n}}{L+1/2}\left\{\frac{\sin^2(\beta_{m,n}^{-}L)}{
\sin(\beta_{m,n}^{-})}+\frac{\sin^2(\beta_{m,n}^{+}L)}{
\sin(\beta_{m,n}^{+})}\right\}
\ee
with $\beta_{m,n}^{\pm}=\pi[m\pm(n-1/2)]/(2L+1)$.

\subsection{Infinite chain limit}

In the infinite system limit, $L \to \infty$, $n/L=O(1)$, $m/L=O(1)$ and $m-n \le \ell$ for
both boundary conditions we recover the known result\cite{lsm,pfeuty}:
\be
G_{m,n}=\frac{2}{\pi} \frac{(-1)^{m-n}}{2(m-n)+1}\;.
\ee

\section{The shift exponent for closed chains}
\label{sec:shift}

To explain the finite-size scaling behavior of $\hat{\cal S}(L,h)$ for
closed chains, we recall the entropy in the infinite system analytically
obtained in Ref.~\cite{peschel05,jin_korepin}.  Here we write it in
terms of the variable, $\gamma=(1-h)/(1+h)$, as:
\beqn
S&=&\frac{1}{12}\left\{ \log_2\left[\frac{4(1-\gamma)^4}{(1+\gamma)^2|\gamma|} \right] \right. \cr
 &+&\left.\frac{2}{\pi \ln 2}
(1+6\gamma+\gamma^2)I(\gamma)I(\gamma') \right\},
\label{S_term}
\eeqn
where $I(\gamma)$ denotes the complete elliptic integral of the first kind and
$\gamma'=\sqrt{1-\gamma^2}$.  The advantage of the form given in
Eq.~(\ref{S_term}) is that it is valid both for $h<1$ and $h>1$.

We note that the entropy is not symmetric  with respect to $h_c=1$; if we
compare its value at $h$ and $h^{-1}$, it is larger in the ordered phase,
$h<1$, by an amount of $\Delta S(h) = S(h)-S(h^{-1})$:
\beqn
\Delta S(h) &=& \frac{1}{2} \left[ \log_2 \frac{1-\gamma}{1+\gamma} +
\frac{4}{\pi \ln 2}I(\gamma)I(\gamma') \right] \cr
& \underset{\gamma\ll 1}{\rightarrow} & \gamma \left[ -\log_2 \gamma +2-\frac{1}{\ln 2} \right].
\eeqn
The last equation for small $\gamma$ is valid in the vicinity of the critical
point. Next we define for each $h<1$ a transverse field $h'>1$ via the relation:
$S(h)=S(h')$. Close to the critical point the distance between $h^{-1}$ and
$h'$ is given by:
\be
\Delta h=h^{-1}-h' \approx \frac{\Delta S(h)}{\partial S/\partial h} \approx
\gamma^2 \left[ -\log_2 \gamma +2-\frac{1}{\ln 2} \right].
\label{Delta_h}
\ee
which vanishes only at the critical point.  Now let us consider a large finite
system of length $L$ at the transverse field $h=h(L)$, where the singularity of
the entropy starts to be rounded (this happens when $[\hat{\cal
S}(L,h)-\hat{\cal S}(\infty,h)]/\hat{\cal S}(\infty,h)$ exceeds some small
limiting value). For closed chains, in which there are the same number of couplings and
transverse fields, the same is true at the corresponding point,
$h'(L)$, too. The position of the maximum of $\hat{\cal S}(L,h)$ is about at
$h_m(L)\approx[h(L)+h'(L)]/2$. If the entropy is symmetric at $h$ and
$h^{-1}$, the estimate of the transition point would be:
$h_c^{\rm{sym}}(L)\approx[h(L)+h^{-1}(L)]/2$, and the distance between
$h_m(L)$ and $h_c^{\rm{sym}}(L)$ is about $\sim \Delta h(L)/2$. This value
is in the same order as the shift of the finite-size transition point. 
Making use of the fact that $L \sim \xi  \sim \gamma^{-1}$, we obtain from
Eq.(\ref{Delta_h})
\be
\Delta h(L) \sim (h_c-h_m(L)) \sim L^{-2} \left[ \log_2 L +a \right]
\label{shift}
\ee
with $a \approx 2-1/\ln 2$. This means that the true value of the shift exponent
is $\lambda=2$, but there is a strong logarithmic correction, which makes the numerical
calculation of $\lambda$ very difficult. 


\end{document}